# The influence of time and discipline on the magnitude of correlations between citation counts and quality scores[1]

Mike Thelwall, Ruth Fairclough
Statistical Cybermetrics Research Group, University of Wolverhampton, UK.

**Abstract**
Although various citation-based indicators are commonly used to help research evaluations, there are ongoing controversies about their value. In response, they are often correlated with quality ratings or with other quantitative indicators in order to partly assess their validity. When correlations are calculated for sets of publications from multiple disciplines or years, however, the magnitude of the correlation coefficient may be reduced, masking the strength of the underlying correlation. In response, this article uses simulations to systematically investigate the extent to which mixing years or disciplines reduces correlations. The results show that mixing two sets of articles with different correlation strengths can reduce the correlation for the combined set to substantially below the average of the two. Moreover, even mixing two sets of articles with the same correlation strength but different mean citation counts can substantially reduce the correlation for the combined set. The extent of the reduction in correlation also depends upon whether the articles assessed have been pre-selected for being high quality and whether the relationship between the quality ratings and citation counts is linear or exponential. The results underline the importance of using homogeneous data sets but also help to interpret correlation coefficients when this is impossible.

## Introduction

The ongoing controversy within the wider scientific community about the use of citation-based indicators in research evaluations underlines the importance of continuing to assess their validity. The most straightforward way is to correlate the citation-based indicators with peer judgements about the value of individual publications. Studies taking this approach have used data from research assessment exercises (Franceschet & Costantini, 2011), surveys of experts (Gottfredson, 1978), the prestige of the publishing journal (Singh, Haddad, & Chow, 2007; Starbuck 2005) or public peer review systems like F1000 (Bornmann & Leydesdorff, 2013; Li & Thelwall, 2012; Mohammadi & Thelwall, 2013; Waltman & Costas, 2014; Wardle, 2010). A statistically significant positive correlation with an independent measure of quality would suggest that the indicators reflect an aspect of quality to some extent. Moreover, it seems intuitively plausible that stronger correlations are likely to reflect stronger relationships between citations and quality. This is not necessarily true, however, because the strength of a correlation always partly reflects the extent to which the two numbers correlated are derived from homogenous sources. In terms of scientometrics, this means that the homogeneity of the set of publications analysed, such as in terms of publication date and subject area, can affect the strength of any correlations derived from it. It is therefore important to understand the impact of a lack of homogeneity on the magnitude of correlation coefficients.

When conducting a scientometric study, the publications analysed can sometimes be classified only crudely by subject area. This mixing may weaken the strength of any relationships between the citation counts and other variables. To illustrate this, suppose that four humanities publications are rated 1, 2, 3, and 4 for quality by a subject expert and receive 1, 2, 3, and 4 citations, respectively. Then there is a perfect relationship between quality and citation counts with a Spearman rank correlation coefficient of 1. Suppose that four life sciences publications are also rated 1, 2, 3, and 4 but, because life sciences articles tend to be more cited, they receive 10, 20, 30, and 40 citations, respectively. For the four life sciences articles there is also a perfect relationship and the correlation coefficient is 1. If the eight articles are collected together then the relationship is no longer perfect and the correlation coefficient is less than half: 0.488 (Figure 1). The mono-disciplinary correlation of 1 for both disciplines is therefore not evident from the mixed set correlation. Thus, anyone calculating a correlation coefficient for a mixed data set could be misled about the underlying monodisciplinary rank correlation strengths. Nevertheless, correlations for combined sets of publications are not necessarily always low. For example, a study of articles rated by peers as part of an Italian research assessment exercise calculated correlations between these ratings and citations to papers published across three years and within quite broad disciplinary areas but still found correlations as strong as 0.8 (for both Physics and Earth Sciences) (Franceschet & Costantini, 2011).

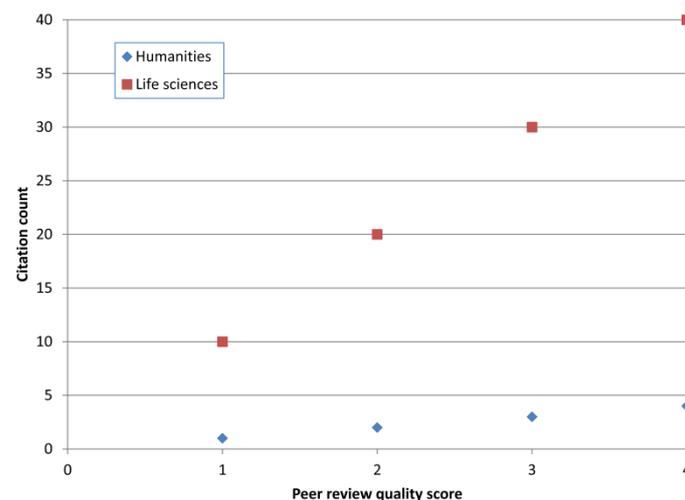

**Figure 1**. Four humanities articles with a perfect linear relationship between quality and citations together with four life sciences articles with a perfect linear relationship between quality and citations. Whilst the mono-disciplinary Spearman correlations are both 1, the combined Spearman correlation is 0.488.

When publications with different citation windows (periods of time for which citations are collected) are merged then this can also reduce the strength of correlation for the same reason as for mixing disciplines. This is because articles with a longer citation window will tend to have higher citation counts than articles with shorter citation windows. As the Figure 1 example shows, combining sets with different average citation counts can substantially reduce the correlation for the mixed set. A previous altmetrics paper has even given an example of a negative correlation between variables (Tweets to papers) despite an underlying positive relationship. The cause of this was that one of the two variables decreased over time (due to a shrinking citation window) and the other increased over time (Thelwall, Haustein, Larivière, & Sugimoto, 2013). The problem of variable citation windows may be unavoidable in some circumstances, however. For example, a study of a narrow group of outputs may need to cover multiple years to get enough data for quantitative analyses. Moreover, combining data from different fields with the same citation window can also be problematic if one attracts citations more quickly than does the other. Although it is clear from a large number of previous scientometric studies that the rate at which articles attract citations varies substantially by field, the cause is not fully understood. The reason could be different citation practices to some extent, but partial coverage of disciplinary outputs by the Web of Science is probably more important (Marx & Bornmann, 2015).

Many studies have also correlated *mean* citation counts for sets of articles with associated peer review scores, such as for the publishing department or institution (Abramo, Cicero, & D'Angelo, 2013; Nederhof & Van Raan, 1993; Oppenheim, 1995, 1997; Oppenheim & Summers, 2008; Rinia, Van Leeuwen, Van Vuren, & Van Raan, 1998; Thomas & Watkins, 1998). This aggregation may also mask the underlying correlation strength but should *increase* it due to the averaging effect of the mean. Peer review judgements have also been correlated with the publishing journal impact factor (Reale, Barbara, & Costantini, 2007), which can indirectly show relationships between citation counts and research quality.

At the abstract mathematical level, it is easy to see that correlations of the same sign can be reduced when data sets with different properties are merged, as the above example shows. In addition, scientometric articles that calculate correlation coefficients sometimes mention this effect when reporting results. Nevertheless, there has been no systematic study of the effects of merging different citation distributions on correlation coefficients. This is a particularly important omission because citation counts tend to follow a highly skewed (Price, 1976) lognormal (Thelwall & Wilson, 2014) distribution and it is therefore not possible to intuitively estimate the amount of masking present in a calculated correlation coefficient. The current article partly fills this gap by assessing the effect of mixing data sets the on the strength of the Spearman correlation coefficient calculated. It uses simulated data with a citation-like distribution to make the estimates and systematically assesses the effect of merging for a range of different plausible options.

## Research Questions

The objective of this study is to assess the degree of masking that occurs when citation counts from multiple disciplines or years are correlated with other data. The following research questions drive the study.

- How much is the correlation between citation counts and another variable reduced when publications from multiple disciplines or years are merged?
- Does the answer to the above question depend upon the properties of the merged sets? The properties considered here are the mean citation scores, the magnitudes of their correlations, when measured separately, the quality distribution of the articles and the relationship between quality, as measured, and citation counts.

## Methods

When data from two (or more) disciplines is merged before a correlation coefficient is calculated, there are two important properties that affect the strength of the correlation of the merged data sets, assuming that the separate data sets follow the same citation distribution, such as the discretised lognormal.

- The mean number of citations per article in each of the two data sets, and, most importantly, the size of the difference between the two means.
- The strength of the correlation between research quality and citation counts in each of the two data sets.

Both of these factors apply when data sets from two different disciplines are merged because the strength of the relationship between citation and research quality could vary between disciplines as well as the average number of citations per paper. Both could also be different for the same discipline across separate years but the most important difference is likely to be that the older articles will have higher citation counts and so the two data sets would have different means. Hence, the overall research design is to test for the effect of merging data sets with differing mean citation counts and differing strengths of the relationship between research quality and citation counts.

The effect of merging data sets was tested for by artificially creating two different data sets with different mean citation counts and different correlations with research quality and then measuring the strength of the correlation with research quality in the merged data set. Since there are many parameters that can be varied, giving a huge number of potential results for a comprehensive evaluation, some of the less important parameters were fixed at the beginning (of the simulations) (but see the Discussion section below): the sample size was set to 5000 for both data sets and the mean of the first data set was fixed at 20, which is a little below the mean number of citations (26.6) reported for articles in the UK Research Excellence Framework (REF) 2014, assuming that the fields with higher average citation rates tended be in the subset for which citation counts were used in the UK REF 2014.

The data sets needed quality scores as well as citation counts for each article. The distribution of different quality scores is likely to vary according to the purpose for which a set of articles is collected. For example, a set of articles from WoS journals is likely to tend to have a higher proportion of highly rated articles than a random set of articles from Google Scholar. Since there is infinite potential variety in this aspect, as a practical step, only two quality distributions were used. The first is the distribution of peer review ratings given to outputs submitted to the UK Research Excellence Framework 2014, calculated from aggregate data reported on its website (results.rae.ac.uk). This gives the distribution: 0: 0.6%; 1*: 3.5%; 2*: 24.0%; 3*: 49.5%; 4*: 22.4%, in ascending order, where 0 corresponds to below national standards of research and 4* corresponds to world-leading research. This is a selective set with a bias towards high quality articles. The relationship between quality, as reflected in REF 2014 scores, and citation counts is probably not linear but closer to exponential, although there is currently no evidence of the exact relationship. Hence the expected means for each quality level were set with ratios 1:2:4:8:16. In other words, for a set of articles the means were fixed at the following values: 0: 1x; 1*: 2x; 2*: 4x; 3*: 8x; 4*: 16x, where x was adjusted to give a pre-selected expected value (mean) of the whole distribution. For comparison purposes, additional tests were conducted with a linear relationship between quality and the citation means (1:2:3:4:5). Tests were also constructed for a non-selective set with an even distribution between the four quality level categories (i.e., 0: 20%; 1*: 20%; 2*: 20%; 3*: 20%; 4*: 20%) since the UK REF 2014 data set is highly selective and many other sets of publications may be far more comprehensive and so may include a larger proportion of lower quality articles.

For both data sets, correlations were varied between 0.1 and 0.9 in increments of 0.1 to give a range of very low to very high plausible values. For the second data set, the mean was varied between 1 and 40 in increments of 1 to give values substantially below and above the mean of the first data set. The range from 1 to 40 was chosen to be a set of common plausible values, although higher means are possible for specialist data sets, such as *Nature* articles from several years ago. Each data set was produced in the statistical package *R* using the *rlnorm* function, which produces a (continuous) lognormal distribution. The numbers were truncated to create a discretised lognormal distribution, after increasing the initial means by 0.5 to partially offset the truncation effect. Truncation was used instead of rounding to integer values because it seems likely that zeros will occur more often in citation data than predicted by a lognormal distribution and truncation is a straightforward way to accommodate this phenomenon.

The *rlnorm* function in *R* is based on the (log of the) mean and standard deviation of the distribution rather than its correlation with another variable and so a simple search loop was built to seek a standard deviation that would give the desired correlation with the simulated research quality scores. For each different set of parameters, 100 simulations were run and the mean of the combined correlations of the 100 results was recorded. In summary, the following fixed and variable parameters were set.

- Both data sets (simultaneously): distribution of quality either selective (mostly high quality publications) or non-selective (even quality spread).
- Both data sets (simultaneously): relationship between quality score and mean citation counts either linear or exponential.
- Data set 1: mean fixed at 20.
- Data set 2: mean varying from 1 to 40 in increments of 1.
- Both data sets (independently): correlation varying from 0.1 to 0.9 in increments of 0.1.
- Both data sets (independently): standard deviation adjusted to give the desired correlation with research quality.
- Both data sets: sample size 5000.
- Experiment: 100 iterations for each unique set of parameters.

## Results

The results are illustrated in separate graphs in which all of the parameters of data set 1 are fixed and all of the parameters of data set 2 vary (Figures 2-13). For example, in Figure 2, the mean of data set 1 is 20 and the data set 1 correlation is 0.1 throughout. Each line in Figure 2 corresponds to a fixed value of the correlation for data set 2, with the height of the line showing the effect of different data set 2 means. Although thirty-six graphs were initially drawn, one for each data set 1 correlation between 0.1 and 0.9 in increments of 0.1 (nine graphs), for each of the four variations of quality distribution and citation-quality relationship (linear or exponential), only the graphs for data set 1 correlations of 0.1, 0.5, and 0.9 are shown since the others are broadly in between these (see http://figshare.com/articles/Time_Discipline_and_Correlation/1299511 for the complete set of 36 graphs and the raw data, and the R software used). Confidence intervals for the means of the correlations (constructed from the 100 iterations) are too small to show in the figures (typical width 0.0001 for 95%).

Figures 2-4 show the results of the experiments with the REF-like selective sets of (mainly high quality) articles and an exponential relationship between quality ratings and citation counts. Unsurprisingly, in all of these graphs the lines for the correlations for merged data sets are always higher when the correlation for the second set is higher. Hence the lines do not cross and higher lines are from higher set 2 correlations. As expected also, the correlation for the merged data set is always identical to the correlation for the two separate data sets when they both have the same mean (i.e., 20) and they both have the same correlation (0.1 in Figure 2, 0.5 in Figure 3 and 0.9 in Figure 4). The fact that the lines are mostly not straight and horizontal also confirms that the citation mean of the two sets affects the strength of correlation for the combined data set. Moreover, the correlation for the combined data sets is always below the average of the correlations of the two separate data sets, except when they are equal.

Beyond the above observations, however, the patterns are more subtle. In Figure 2, when data set 2 has a mean citation count above 5, the combined correlation is less than a third of the average of the correlations of the two data sets. For example, when the second data set correlation is 0.9 (the highest line in Figure 2) and the set 2 mean is 20, the combined data set has a correlation of about 0.28, which is below the average (0.5) of the two data set correlations (0.1, 0.9). When the data set 2 mean is 1, however, the combined correlation is much higher at 0.44. This pattern almost reverses in Figures 3 and 4, however, but Figures 8 and 11, both with data set 1 correlation equal to 0.1, have similar shapes.

The shape of Figure 3 is more typical than the shape of Figure 2 in the sense that a broadly similar shape is evident in most of the other graphs and the exceptions, with very low correlations, may be unusual cases in practice. In Figure 3, as in most of the graphs, the correlation for the mixed distribution is lowest when the citation means of the two individual distributions differ most, reflecting the Figure 1 example. The difference is strongest when the ratio between the two means is highest (i.e., when the set 2 mean is close to 1).

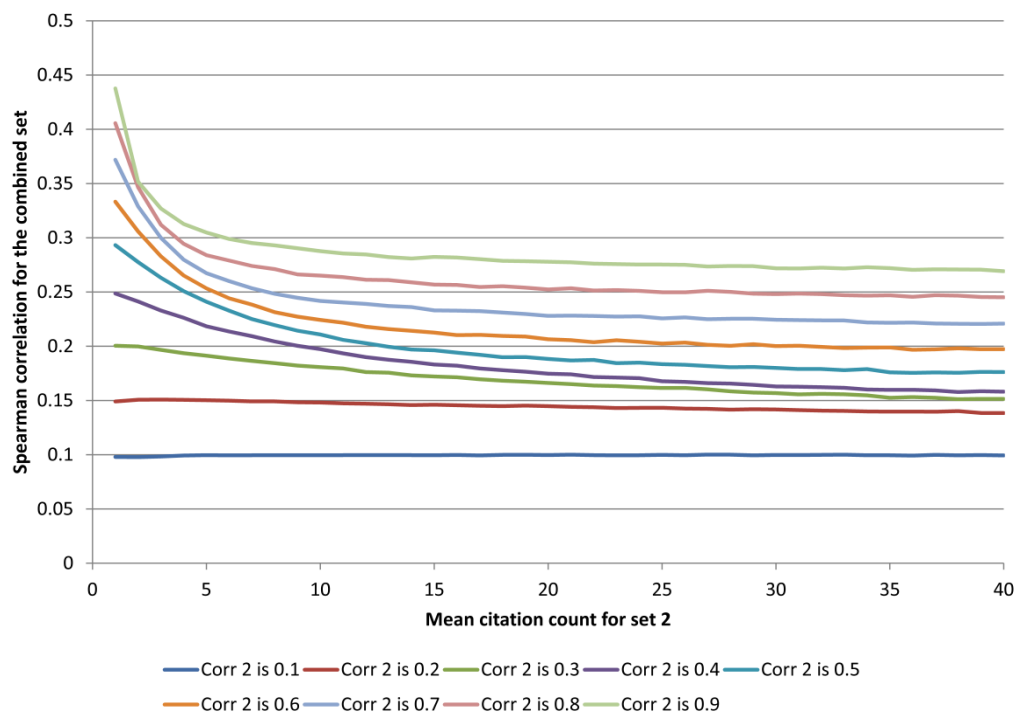

Figure 2. Spearman correlations between research quality and citation counts for two simulated data sets of 5000 papers, the first having a citation mean of 20 and a Spearman correlation of 0.1 between research quality and citation counts (selected quality, exponential citation relationship).

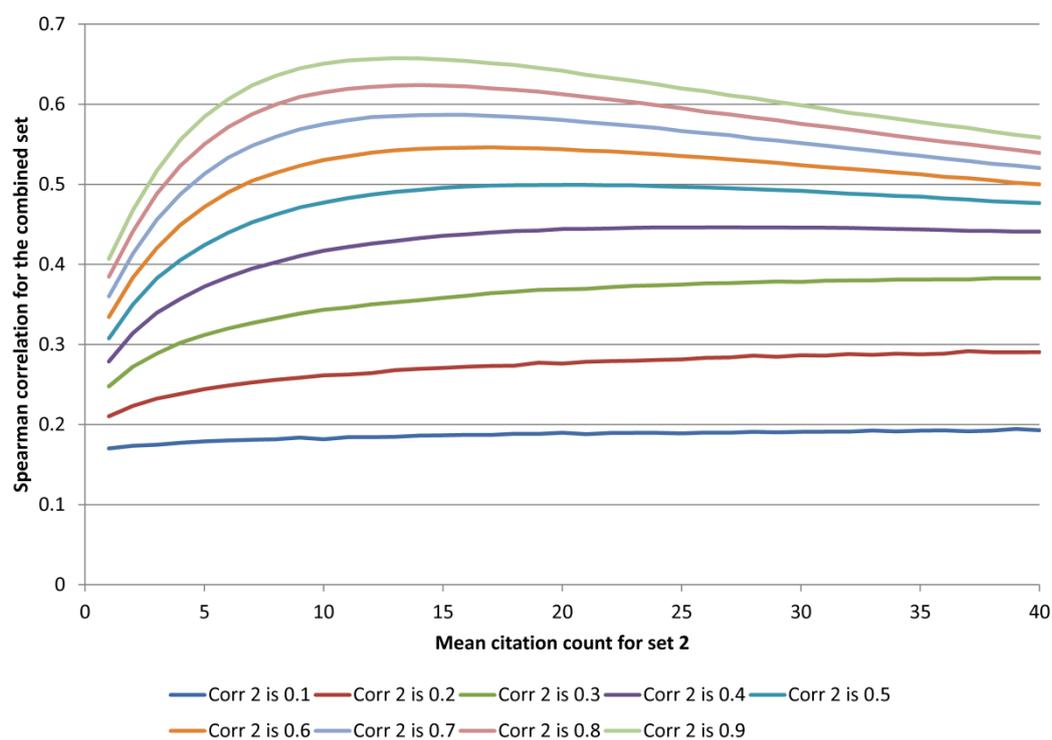

Figure 3. Spearman correlations between research quality and citation counts for two simulated data sets of 5000 papers, the first having a citation mean of 20 and a Spearman correlation of 0.5 between research quality and citation counts (selected quality, exponential citation relationship).

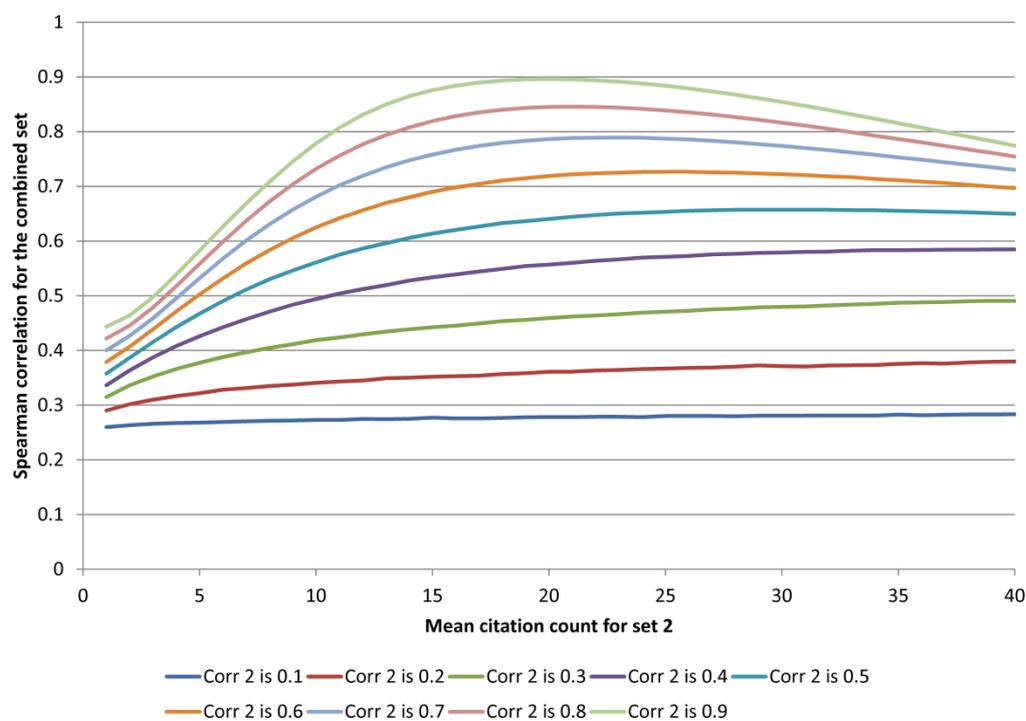

Figure 4. Spearman correlations between research quality and citation counts for two simulated data sets of 5000 papers, the first having a citation mean of 20 and a Spearman

correlation of 0.9 between research quality and citation counts (selected quality, exponential citation relationship).

In Figures 5-7, where the relationship between quality and citation counts is linear rather than exponential, the typical graph (Figure 6) is much more sensitive to differences in the mean. Hence, the bell-shaped graph reflects a reasonably dramatic fall in the correlation for a mixed set of articles with different correlations when the means of the two sets are not the same. So, for example, sets of articles published in two consecutive years might easily have citation means of 20 and 25, but merging them would reduce the overall correlation substantially. If their individual correlations were 0.5 and 0.9, respectively, then according to Figure 5, their combined correlation would be only 0.6.

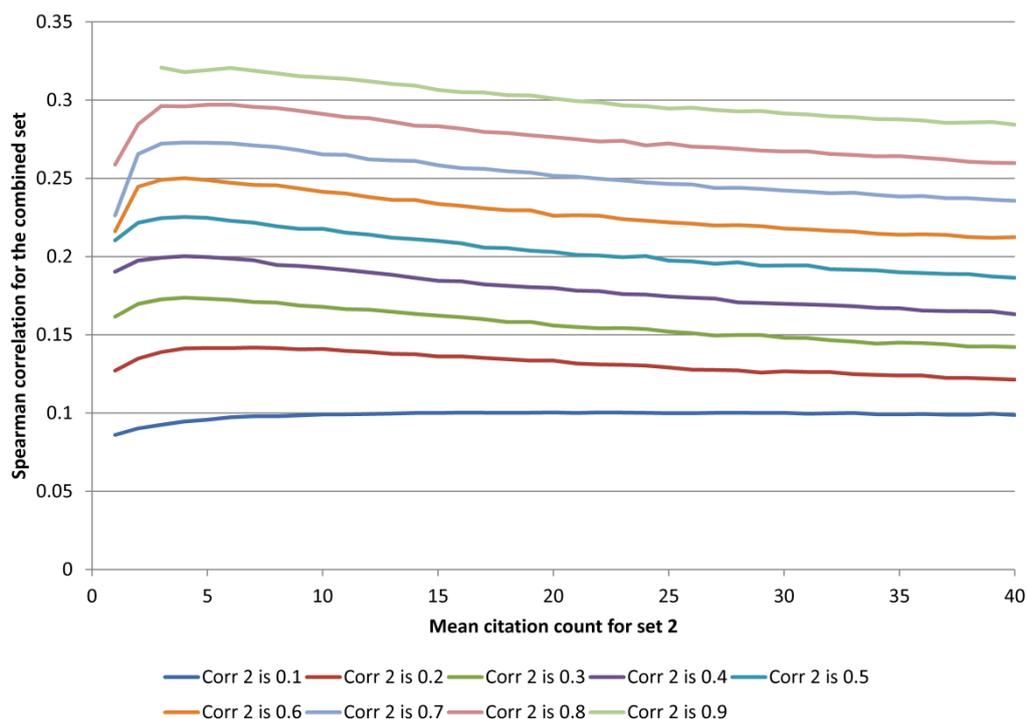

Figure 5. Spearman correlations between research quality and citation counts for two simulated data sets of 5000 papers, the first having a citation mean of 20 and a Spearman correlation of 0.1 between research quality and citation counts (selected quality, linear citation relationship).

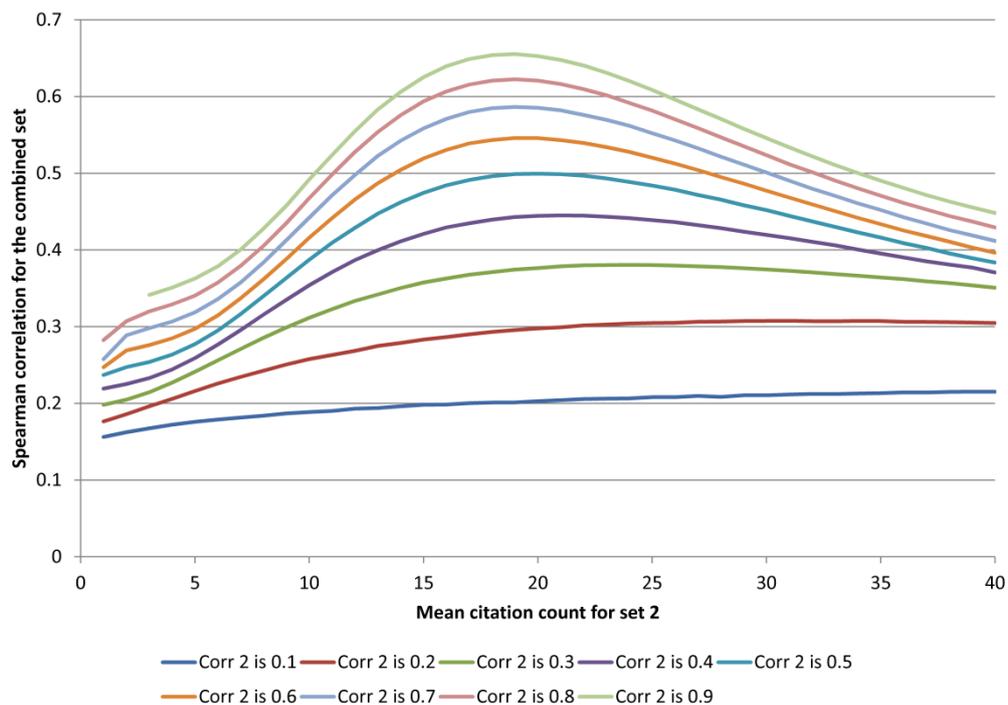

Figure 6. Spearman correlations between research quality and citation counts for two simulated data sets of 5000 papers, the first having a citation mean of 20 and a Spearman correlation of 0.5 between research quality and citation counts (selected quality, linear citation relationship).

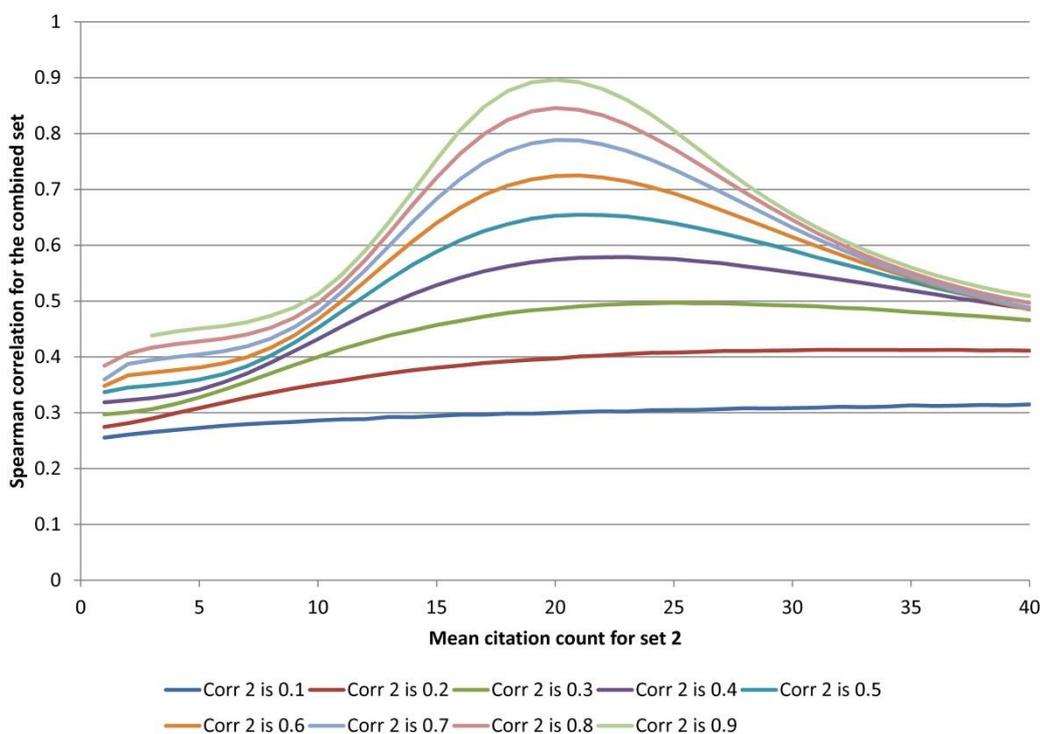

Figure 7. Spearman correlations between research quality and citation counts for two simulated data sets of 5000 papers, the first having a citation mean of 20 and a Spearman correlation of 0.9 between research quality and citation counts (selected quality, linear citation relationship).

In Figures 8-10, there is an exponential relationship between quality and citation counts but the sample of articles is non-selective (i.e., equal numbers of articles from each quality level). In these graphs, the lines are relatively flat compared to the others, and particularly in the case of Figure 9.

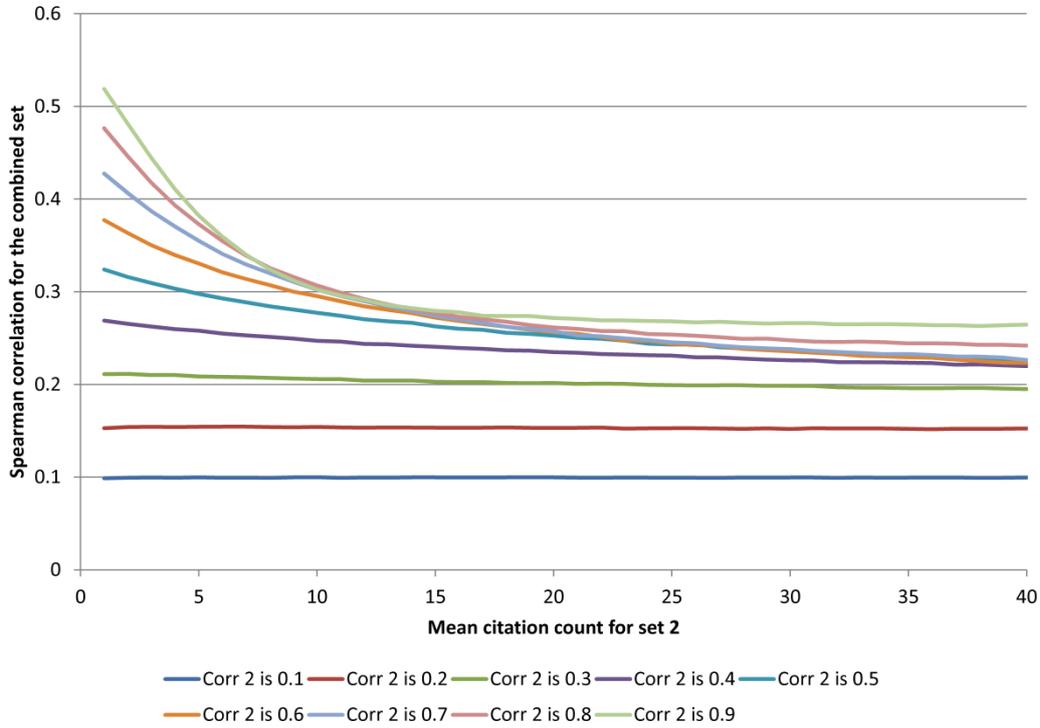

Figure 8. Spearman correlations between research quality and citation counts for two simulated data sets of 5000 papers, the first having a citation mean of 20 and a Spearman correlation of 0.1 between research quality and citation counts (non-selected quality, exponential citation relationship).

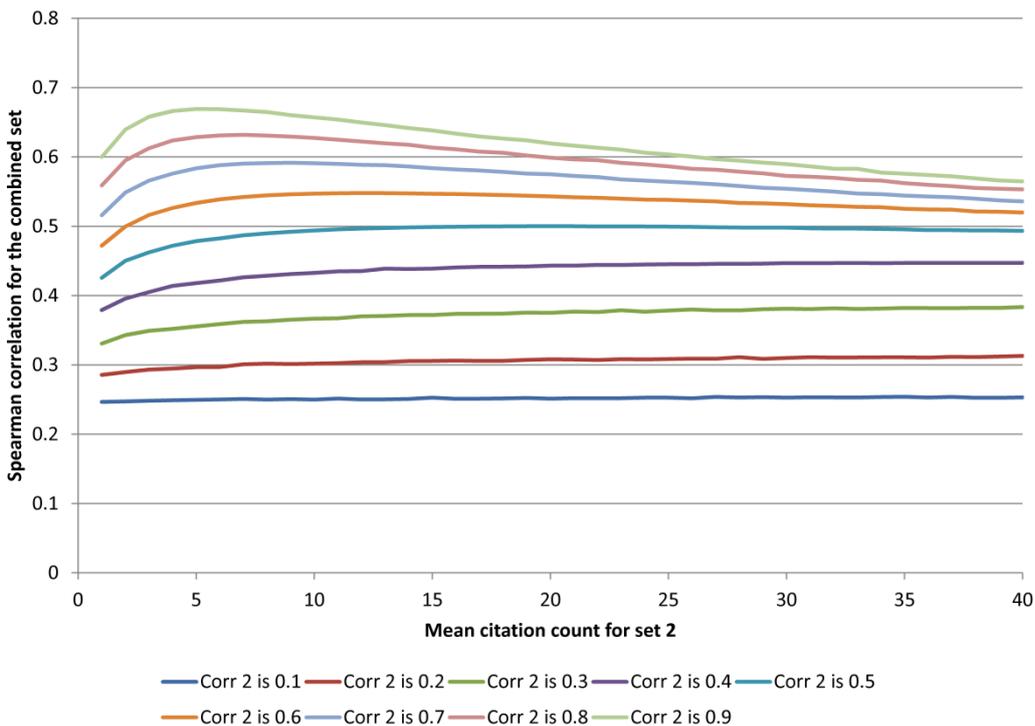

Figure 9. Spearman correlations between research quality and citation counts for two simulated data sets of 5000 papers, the first having a citation mean of 20 and a Spearman correlation of 0.5 between research quality and citation counts (non-selected quality, exponential citation relationship).

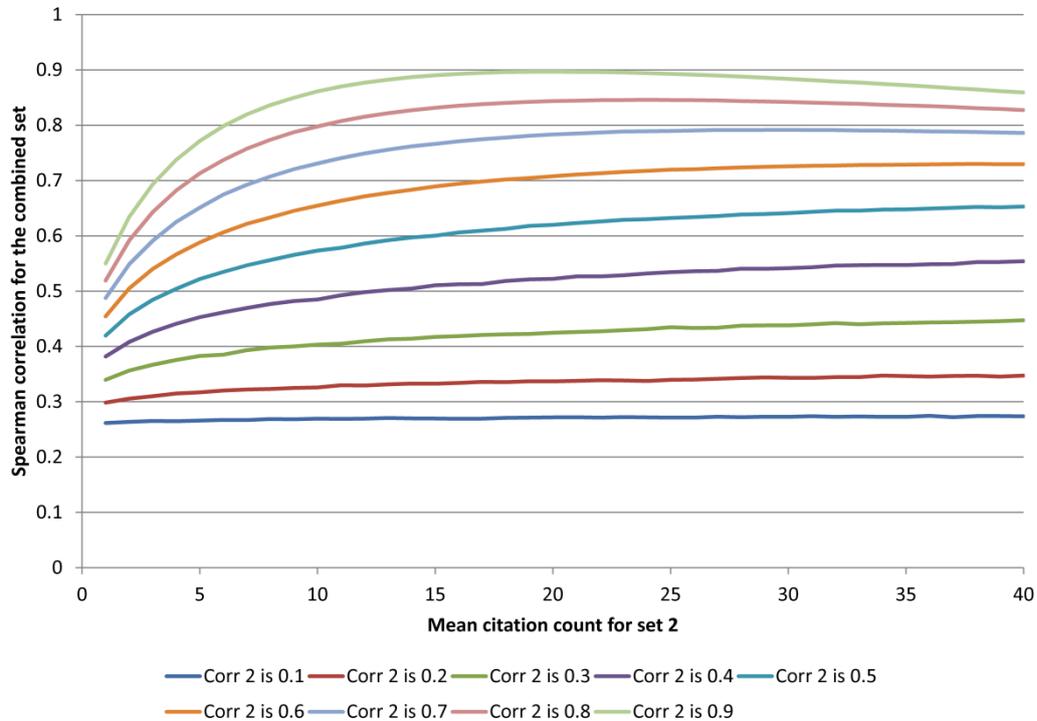

Figure 10. Spearman correlations between research quality and citation counts for two simulated data sets of 5000 papers, the first having a citation mean of 20 and a Spearman correlation of 0.9 between research quality and citation counts (non-selected quality, exponential citation relationship).

Figures 11-13 show the case where the article sets are non-selective and the relationship between quality and citation counts is linear. These three graphs are almost identical to Figures 2-4, despite the different assumptions behind them (both experiments were also repeated three times to make sure that this was not an experimental error), and so the same comments apply.

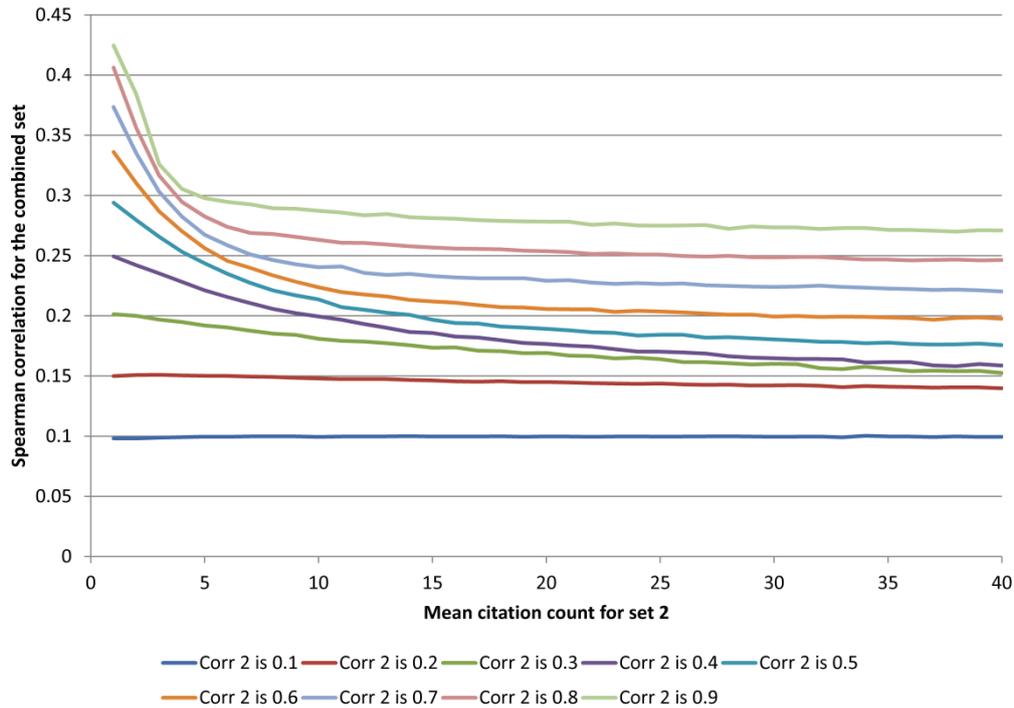

Figure 11. Spearman correlations between research quality and citation counts for two simulated data sets of 5000 papers, the first having a citation mean of 20 and a Spearman correlation of 0.1 between research quality and citation counts (non-selected quality, linear citation relationship).

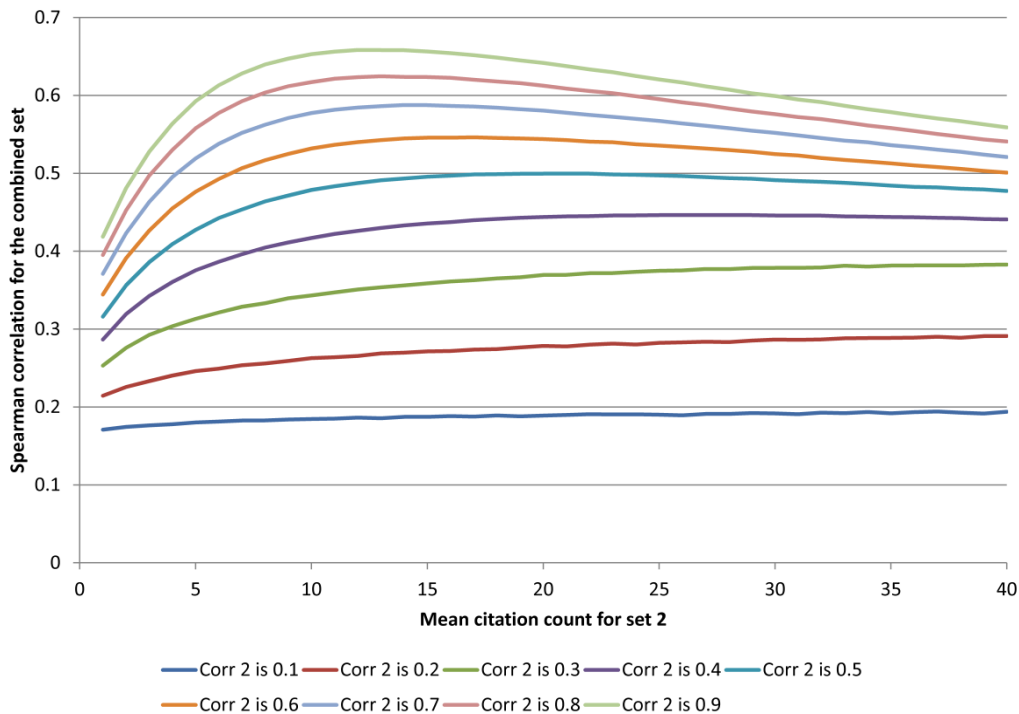

Figure 12. Spearman correlations between research quality and citation counts for two simulated data sets of 5000 papers, the first having a citation mean of 20 and a Spearman correlation of 0.5 between research quality and citation counts (non-selected quality, linear citation relationship).

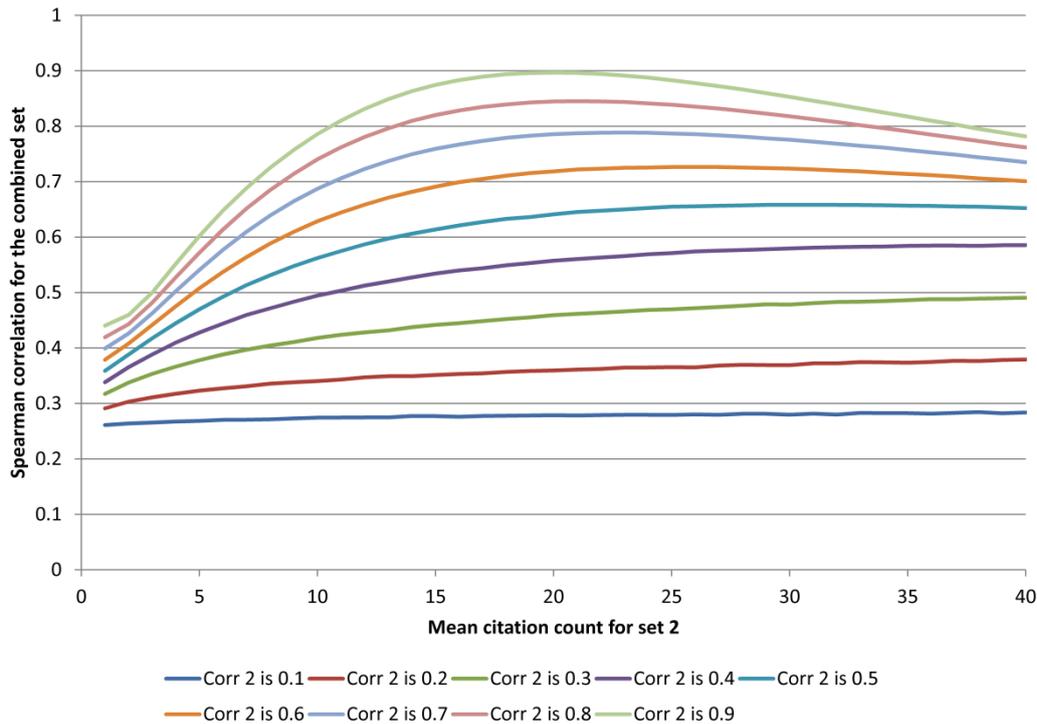

Figure 13. Spearman correlations between research quality and citation counts for two simulated data sets of 5000 papers, the first having a citation mean of 20 and a Spearman correlation of 0.9 between research quality and citation counts (non-selected quality, linear citation relationship).

When merging datasets with different means the key factor is the ratio between the two means. For example, when two data sets have the same correlation but one has double the mean of the other, then the combined data set correlation should be approximately the same whatever the magnitude of the means. For example, in figures 2-13 above, the height of the line for set 1 and set 2 having equal correlations is the same for the set 2 mean 40 as for the set 2 mean 10 since in both cases one set mean is double the other (20:40 or 20:10). When the correlations in the two data sets are not the same, however, this is not true. For example, the line representing a data set 2 correlation of 0.9 in Figure 3 (where the set 1 correlation is 0.5) is higher at a set 1 mean of 10 than at a set 1 mean of 40, despite both representing a doubling of the means in comparison to the mean of 20 for set 2.

An important pattern is that the correlation 0.1 line is relatively stable (flat) in most figures. This shows that the magnitude of the difference in means rarely makes much difference to the correlation for the combined set when one of the two correlations is close to zero. Figures 5-7 for low values of the set 2 mean and Figure 3 to a lesser extent are the main exceptions to this rule.

## Discussion

The results discussed above have a number of limitations that prevent them from comprehensively answering the research questions. The shapes of the graphs may be different for data sets with different mean citation levels, such as if both data sets had very low mean citation counts. Moreover, in practice, some combined data sets may merge 3 or more coherent subsets and the impact of this on the correlation has not been assessed. Perhaps more fundamentally, it seems likely that no collection of articles will have a citation distribution that is pure (i.e., homogeneous in terms publication dates, subject areas and all possible other factors) but that it will merge a number of similar distributions each of which are affected to some extent by a range of factors, such as author nationality (Didegah & Thelwall, 2013). Moreover, citation windows are only approximations because articles are published at different times during the year, so a citation distribution for articles from a single year would aggregate a continuously changing theoretical daily set of citation distributions (Levitt & Thelwall, 2011; similar to: Burrell, 2007). Perhaps most fundamentally, science is interconnected and so there are probably few areas of research that are capable of generating sets of articles with homogenous citation distributions.

Figure 14 shows two extreme distributions and the effect of merging them into a single data set. As can be seen in the online colour version, the points for the high correlation set form solid blocks with little variety at each quality rating, whereas the points for the low correlation set include many zeros at all quality ratings, but more very high citation counts at the higher citation ratings. As a result of the high spread for the low correlation data set, most of the highest (5) rated articles are uncited for this set. Thus, the large number of highly rated articles with the lowest possible score is the cause of the overall dramatic reduction in correlation for the combined set. More generally, the uncited articles in data sets that do not have a high correlation between quality and citation counts are presumably the underlying reason why the correlation for a combined data set is below the mean of the correlations of the two individual data sets.

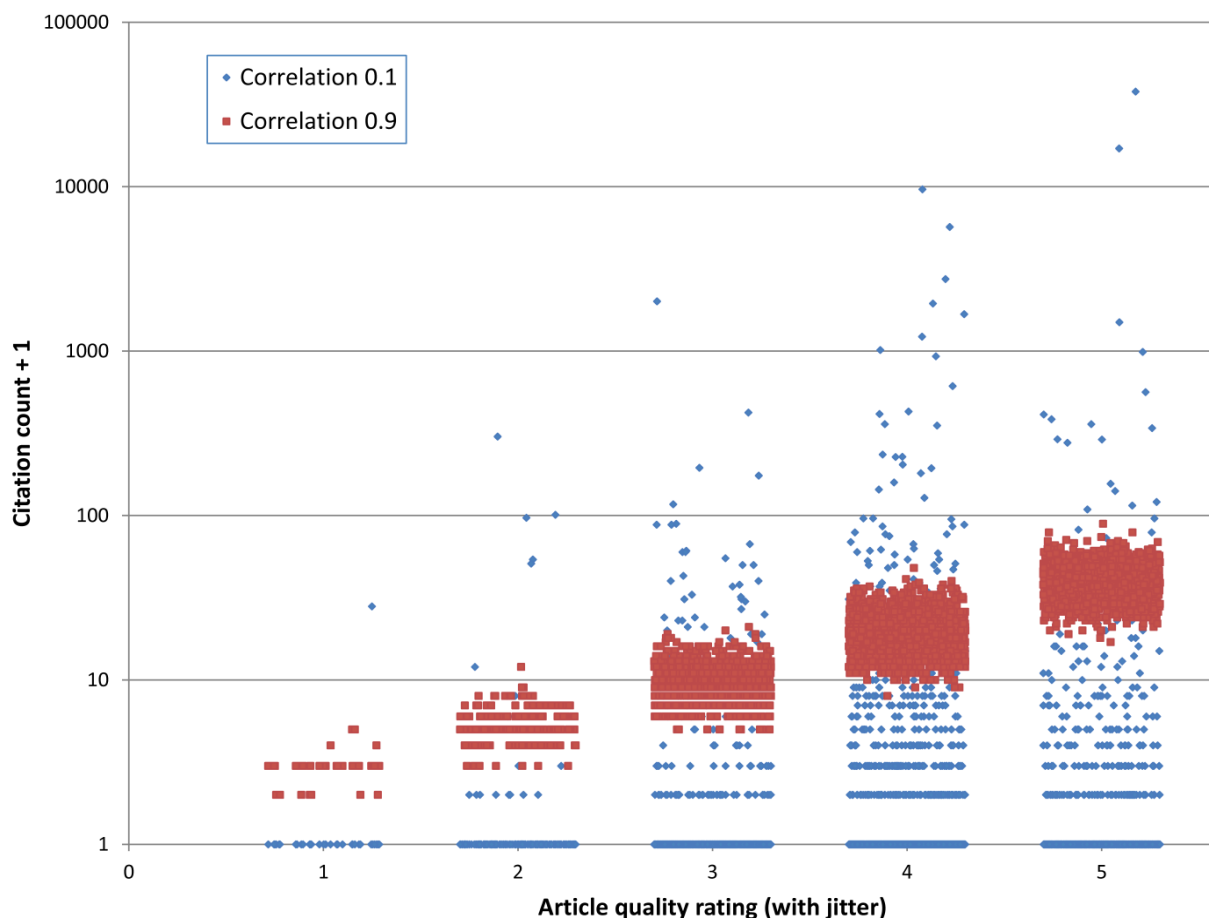

Figure 14. Two simulated citation data sets with mean 20 and correlations of 0.1 and 0.9 with quality (selective article sets, exponential quality-citations relationship, n=5000 each). Note the logarithmic scale on the y axis and that 1 is added to citation counts to allow zeros

to be plotted on the log scale. Random jitter has been added to the quality scores to allow the graph to indicate the number of duplicate points at given values.

In psychology, guidelines are sometimes given about how to interpret the magnitude of correlation coefficients. The advice typically takes the form of recommending specific terms, such as weak, medium, or strong, for specific ranges of values. This is made on the basis of the typical range of values reported in the literature in the belief that a perfect correlation is impossible due to natural variations in people's behaviours and so judgements should be made by comparing values of the correlation coefficient calculated in real studies (Hemphill, 2003). As the experiments above show, this approach will not work in scientometrics without taking into account the effect of heterogeneity in the data sets examined. Hence, if future studies can use reasonably homogenous sets of papers then direct comparisons between results would give convincing evidence about the strength of the relationship between research quality and citation counts in different fields. In cases where a study is forced to analyse a heterogeneous population, such as with a variable citation window or with multiple fields grouped together, then a partial solution would be to estimate the key parameters of the constituent distributions and use this information to estimate the strengths of the correlations for the homogenous subsets from the overall correlation. Figures 2-13 could be used to compare the measured overall correlation with the hypothesised parameters of the constituent subsets, assuming that there were two of them. For this, however, the means of the different subsets would be needed as well as initial estimates of the likely correlation strengths, the distribution of article quality (e.g., selective or non-selective) and the relationship between citation counts and the quality metric (e.g., linear or exponential). The R code associated with this article can be modified for this purpose, or the solution may be estimated from the graphs above.

An alternative solution to the problem of merging distributions with different means is to use field-normalised citation counts so that all sets of articles come from a distribution with mean 1. When this field normalisation is possible and is relatively fine-grained it should virtually eliminate the problem of merging data sets with different means but not the problem of merging data sets with different correlations between (normalised) citation counts and quality scores.

## Conclusions

The results show that correlations between research quality and citation counts are unreliable when calculated for heterogeneous sets of articles, such as those with variable citation windows (e.g., from multiple years) or from multiple fields. In particular, the combined correlation can be expected to be substantially below the mean of the correlations for the constituent subsets, at least in the case of two equally-sized subsets. This appears to be due to the relatively high number of uncited articles, even at high quality levels, unless there is a very high correlation between quality and citation counts.

The extent to which the combined correlation falls below the mean for the two sets varies considerably but tends to be larger when the means of the two distributions are different. It also varies according to the nature of the sample analysed (e.g., selective or non-selective) and the relationship between the quality metric and citation counts (e.g., linear or exponential). Hence, estimating the strength of the underlying correlations is not straightforward when only a combined correlation is available. In such circumstances, the best available option is to estimate the key parameters of the constituent homogeneous subsets and to simulate the effect of merging them on the correlation for the combined set.

More generally, for research evaluation purposes it is important to recognise that the extent to which the magnitude of a correlation coefficient reflects the level of the underlying correlation is low for heterogeneous sets of articles. Coefficients should therefore only be interpreted together with contextual information about the homogeneity of the set of publications analysed. In particular, low correlation coefficients could reflect the lack of an underlying correlation or the heterogeneity of the set of articles analysed. This seems to be particularly important for heterogeneous research areas that are often analysed together as a combined set, such as the humanities.